\documentclass[12pt]{article}
 \usepackage[dvips]{graphicx}
\begin{document}
\title {Fuzziness in Quantum Mechanics}
\bigskip
\author{~ Alex Granik\thanks{Department of Physics, University of the Pacific,
Stockton,CA.95211;~E-mail:agranik@uop.edu}~~and~~
H.J.Caulfield\thanks{Fisk University 1000 17th Ave., N.Nashville,
TN 37208 }}
\date{}
\maketitle

\begin{abstract}
It is shown that quantum mechanics can be regarded as what one
might call a "fuzzy" mechanics whose underlying logic is the
fuzzy one, in contradistinction to  the classical "crisp" logic.
Therefore classical mechanics can be viewed as a crisp limit of a
"fuzzy" quantum mechanics. Based on these considerations it is
possible to arrive at the Schroedinger equation directly from the
Hamilton-Jacobi equation. The link between these equations is
based on the fact that a unique ( "crisp") trajectory of a
classical particle emerges out of a continuum of possible paths
collapsing to a single trajectory according to the principle of
least action. This can be interpreted as a consequence of an
assumption that a quantum "particle" "resides" in every path of
the continuum of paths which collapse to a single( unique)
trajectory of an observed classical motion. A wave function then
is treated as a function describing a {deterministic} entity
having a fuzzy character. As a consequence of such an
interpretation, the complimentarity principle and wave-particle
duality can be abandoned in favor of a fuzzy {deterministic}
microoobject.
\end{abstract}
\section{Introduction}
One of the purposes of this paper is to bring together fuzzy
logic and quantum mechanics. Here we extend our prior analysis
\cite{GC1} of a fuzzy logic interpretation of quantum mechanics
by demonstrating that the Schroedinger equation can be deduced
from the assumptions of the fuzziness underlying not only quantum
but also classical mechanics. A pedestrian way of defining
fuzziness was given by Kosko \cite{BC1} who wrote that the fuzzy
principle states that everything is a matter of degree. More
rigorously, fuzziness can be defined as multivalence.\\

Interestingly enough, even separation between classical and
quantum domains is somewhat fuzzy since there is no crisp
boundary separating them ( see, for example, \cite{IP}).
Moreover, we can even claim that the difference between these
domains is only in a degree of fuzziness. In fact, both classical
and quantum mechanics make predictions based on repetitive
measurements which imply a certain spread of results.\\

The crisp character of the formal apparatus of classical
mechanics hides this important fact by a seemingly absolute
character of a single measurement. From this point of view the
ultimate statements of classical mechanics are nothing but the
results a certain averaging  ( or defuzzification, meaning the
elimination of the spread) with some weight which we call the
"fuzziness density". The latter can be represented by some
function. The fuzziness density then varies from "sharp' ( in
classical mechanics) to "diffuse" (in quantum mechanics).\\

If we consider a concept of a "thing in itself" and assume(quite
plausibly)  that it has a fuzzy and {deterministic} character,
then in  a series of experiments designed to elicit its properties
to the outside observers it appears as a random set thus
disguising its deterministic nature. As we have already
indicated, we consider classical and quantum mechanics as having
common fuzzy roots and no sharp dividing boundary. They can be
viewed as different realizations of a fuzzy "thing in itself".
This can explain why some phenomena in a strongly fuzzy domain of
quantum mechanics cannot be realized in a weakly fuzzy (more
precisely, zero fuzzy) domain of classical mechanics.\\

Thus if we accept quantum mechanics as a more general theory than
classical mechanics, then it seems reasonable to expect that the
former could  be constructed independently from the latter.
However the basic postulates of quantum mechanics cannot be
formulated , even in principle, without invoking some concepts of
classical mechanics. Both theories share some basic common
feature, namely that they are rooted in the fuzzy reality. This
somehow justifies a paradoxical statement by Goldstein (as quoted
in \cite{MK}) that quantum mechanics is a repetition of classical
mechanics suitably understood.\\

Our basic assumption is that reality is fuzzy and nonlocal not
only in space but also in time. In this sense idealized pointlike
particles of classical mechanics corresponding to the ultimate
"sharpness" of the fuzziness density emerge  in a process of
interaction between different parts of fuzzy wholeness. This
process is viewed as a continuous process of defuzzification. It
transforms a fuzzy reality into a crisp one. It is clear that the
emerging crisp reality (understood as a final step of
measurements which we call detection) carries less information
that the underlying fuzzy reality. This means that there is an
irreversible loss loss of information usually called a collapse
of the wave function within a context of quantum mechanics. From
our point of view it is not so much a "collapse' as a realization
of one of many possibilities existing within  a fuzzy reality. Any
measurements (viewed as a process) rearranges the fuzzy reality
leading to different detection outcomes according to the changed
fuzziness.\\

Therefore it seems quite reasonable to expect that classical
theory bears some traces of quantum theory underlying (and
connected to)it. In view of this we would like to recall the
words  by Bridgeman who remarked  that the seeds and the sources
of the ineptness of our thinking in the microscopic range are
already contained in our present thinking applied to a large-scale
regions. One should have been capable of discovery of the former
by a sufficiently critical analysis of our ordinary common sense
thinking.
\section{Some Basic Concepts}
As we have already indicated, both classical and quantum
mechanics can be viewed as statistical theories (cf.
\cite{LM})with respect to an ensemble of repetitive measurements
where each measurement must be carried out under the identical
conditions. The latter is a very restrictive requirement dictated
by a crisp-logical world view and therefore not attainable even
in a more general setting of fuzzy reality. On the other hand, if
we assume a fuzzy nature of "things" then the apparent statistical
character of physical phenomena would follow not from their
intrinsic randomness but from their fuzzy-deterministic nature.
Outwardly the latter expresses itself as randomness. Clearly,
this definition of the apparent statistical nature of classical
and quantum mechanics is applicable even to one measurement.

Let us elaborate on this.Conventionally, statistical theories are
tied to randomness. However recent results in the theory of fuzzy
logic provided a deterministic definition of the relative
frequency count of identical outcomes by expressing it as a
measure of a subsethood $S(A,B)$, that is a degree to which a set
$A$ is a subset of a set $B$ \cite{BK2}. To make it more concrete
suppose that set $B$ contains $N$ trials and set $A$ contains
$N_A$ $\it succesful$ trials. Then $S(A,B)=N_A/N$.\\

We would like to extend this concept to experimental outcomes of
measurements performed on a classical particle. This would be
possible if we were to to consider the classical particle to be
located simultaneously on all possible paths connecting two
spatial points. In a sense it is not so far fetched since it is
analogous to the idea used by the least action principle.\\

To adapt the concept of fuzziness to a spatial localization of a
particle we introduce the notion of the particle's membership in
a spatial interval ( one-, two, or three-dimensional). This
membership, generally speaking, is going to vary from one
interval to another. We define the membership as follows. Let us
say that we perform $N$ measurements aimed at detecting the
particle in a certain spatial interval. It turns out that the
particle is found in this interval $N_A$ times. The membership of
the particle in the interval is then defined as $N_A/N$ and can
be formally described with the help of Zadeh \cite{Z1}
sigma-function.\\

As a next step, this approach allows us to formally introduce the
{\it {membership ~density}} defined as the derivative of the
membership function. If we denote the membership density by
$\mu$, then a degree of membership of the particle say in an
elemental volume $\Delta V$ is $\mu \Delta V$. According to this
definition the particle has a $zero$ membership in a spatial
interval of measure $0$, that is at a point. Such an apparently
paradoxical result indicates that in general we should base our
estimation of fuzziness on the relative degree of membership
instead of the absolute one.\\

In other words, given a degree of membership $\mu(x_i) dV$ of a
particle in a volume $dV$ containing the point  $x_i$ and a
degree of membership $\mu(x_j) dV$ of the same particle in a
volume $dV$ containing the point $x_j$, we find the $relative
~~degree$ of membership of the particle in both volumes:
$\mu(x_i)/\mu(x_j)$. The same expression represents also the
relative degree of membership of the particle in two points $x_i$
and $x_j$ despite the fact that the absolute degree of membership
of the particle in either point is $0$.\\

An importance of the relative degree of membership is due to the
fact that experimentally a location of the particle is evaluated
on the basis of its detection at a certain location in $N_i$
experimental trials out of $N$ trials. As was shown by Kosko
\cite{BK2}, the ratio $N_i/N$ then measures the degree to which  a
sample of all elementary outcomes of the experiments is a subset
of a space of the successful outcomes. In other words, this ratio
represents a degree of membership of the sample space in the
space of the successful outcomes. In our case the relative degree
of  membership $\mu(x_i)/\mu(x_j)$ of a particle in two points
can be identified as the relative count of the successful outcomes
(in a series of measurements) of finding the particle at points
$x_i$ and $x_j$.\\

In view of these definitions the classical mechanical sigma-curve
of particle's membership in a spatial interval is nothing but a
step function. This simply means that up to a certain spatial
point $x$ the degree of particle's membership in an interval
($-\infty,x]$ is $0$, and for any value $y > x$ the degree of
particle's membership in the interval ($x,y]$ is $1$. The
corresponding membership density is the delta function. Thus the
idealized picture of classical-mechanical phenomena with particles
occupying intervals of measure zero corresponds to the statement
that these particles are $\it strictly$ non-fuzzy , their
behavior is governed by a crisp bivalent logic, and the
respective membership density is the delta-function.\\

In reality, any physical "particle" occupies a small but nonzero
spatial interval. This means that the membership density is a
sharp ( but not delta-like) function corresponding to a minimum
fuzziness. At  the other end of the spectrum, in the microworld,
the fuzziness is maximal. In fact, if we accept the idea that a
quantum mechanical "particle" (a microobject) "resides" in
different elemental volumes $dV$ of a three-dimensional space
with the varying degrees of residence (membership), then we can
apply to such a microobject our concept of the membership
density. In general, this density cannot be made arbitrarily
narrow as is the case for a classical particle. The latter can be
considered as the limiting case of the former when the membership
density becomes delta-function-like. Moreover, the fuzziness in
the microworld is even more subtle since mathematically it is
described with the help of complex-valued functions.\\

The latter results in the emergence of the interference phenomenon
for microobjects, which in the classical domain is an exclusive
property of waves. Therefore, mutually exclusive concepts of
particles and waves in classical mechanics become inapplicable in
the realm of fuzzy reality where "particles" and "waves" are not
mutually exclusive concepts, but rather different expressions of
fuzziness. For example, the double-slit experiment can be
interpreted now as a microobject's "interference with itself"
because it has a simultaneous membership in all parts of space
including elemental volumes containing both slits. Since the total
membership of a microobject in a given finite volume is fixed, any
change of its membership in one of the slits affects the
membership everywhere leading to the interference effects.\\

In the following we "recover" the fuzziness of the quantum world
by deriving the Schroedinger equation from the Hamilton-Jacobi
equation, where the latter can be viewed as the result of the
 fuzziness reduction ( destruction) of the quantum world.

\section{DERIVATION OF THE SCHROEDINGER EQUATION}

First, we show how the Hamilton-Jacobi equation for a classical
particle in a conservative field can be derived from Newton's
second law, thus connecting it to the destruction of fuzziness.
In principle, a particle's motion between two fixed points, A and
B, can  occur along any conceivable path (a "fuzzy" ensemble in a
sense that a particle has membership in each of the paths)
connecting these two points. In the observable reality these paths
"collapse" onto one observable path. Mathematically, this
reduction is achieved by imposing a certain restriction on a
certain global quantity (the action $S$), defined on the above
family of paths.\\

Let us consider Newton's second law and assume that trajectories
connecting points $A$ and $B$ comprise a continuous set. This
means in particular that the classical velocity is now a function
of both the time and space coordinates, $\bf v =\bf v(\bf
x$,$t)$. Now we fix time $t =t_0$. Since on the above set for the
fixed $t_0$ the correspondence $\bf x$ to $t$ is many to one,
$\bf x$ is not fixed (as was the case for a single trajectory),
and therefore the velocity would vary with $\bf x$. Physically
this is equivalent to considering points on different
trajectories at the same time. Our assumption means that now we
must use the total time derivative:
\begin{equation}
d/dt =\partial /\partial t + \bf V \bullet \bf \nabla
\end{equation}

By applying the curl operation to Newton's second law for a single
particle and performing elementary vector operations we obtain
\begin{equation}
(\partial/\partial t) {\bf\nabla \times p} - (1/m){\bf \nabla
\times (p \times\nabla\times p)} = 0,
\end{equation}
where $p = m  {\bf v}$ is the particle's momentum. If we view (2)
as the equation with respect to $\bf {\nabla}\times \bf{p}$, then
one of its solutions is
\begin{equation}
\bf{p}=\nabla S
\end{equation}
where $S(\bf{x},t)$ is some scalar function to be found. Since
$S$ is defined on the continuum of paths it can serve as a
function related to the notion of fuzziness (here a continuum of
possible paths).\\

Note that the spatial and time variables enter into $S$ on equal
footing. Upon substitution of (2) back in Newton's second law, $d
\bf{p}/dt = -\nabla V$, where $d/dt$ is understood in the sense
of (1), we obtain $$ \nabla [\partial S/\partial t +
(l/2m)(\nabla S)^2 + V] = 0$$.

Integrating this equation and incorporating the constant of
integration (which, generally speaking, is some function of time)
into the function $S$, we arrive at the determining equation for
the function $S$ which is the familiar Hamilton-Jacobi equation
for a classical particle in a potential field $V$:
\begin{equation}
\partial S/\partial t + (l/2m)(\nabla S)^2 + V = 0.
\end{equation}
By using Eqs. (1) and (3) we can represent $S$ as a functional
defined on the continuum of paths connecting two given points,
say $0$ and $1$, corresponding to the moments of time $t_0$ and
$t_1$. To this end we rewrite (4):
\begin{equation}
dS/dt = p^2/2m +V
\end{equation}
Integrating (5) we obtain the explicit expression of $S$ in the
form of the following functional:
\begin{equation}
S=\int_{t_0}^{t_1}(\frac{p^2}{2m}-V)dt,
\end{equation}
which is the well-known definition of the action for a particle
moving in the potential field $V$. Thus we have connected the
concept of fuzziness in classical mechanics with the action $S$.
If we consider $S$ as a measure of fuzziness in accordance with
our previous discussion, then by minimizing this functional
(i.e., by postulating the principle of least action) we
"eliminate" (or rather minimize) fuzziness by generating the
unique trajectory of a classical particle. In a certain sense the
principle of least action serves as a defuzzification procedure.\\

Now we proceed with the derivation of the Schroedinger equation.
There are two basic experimental facts that make microobjects so
different from classical particles. First, all the microscale
phenomena are linear. Second (which is a corollary of the first),
these phenomena obey the superposition principle. Here it would
be useful to recall that even at the initial stages of
development of quantum mechanics Dirac formulated its fuzzy
character, albeit without using the modern-day terminology. He
wrote: " ... whenever the system is definitely in one state we
can consider it as being partly in each of two or more other
states"\cite{PD}. This is as close as one can come to the concepts
of fuzzy sets and subsethood \cite{BK3} without directly
formulating them. In view of this it does not seem strange that a
microobject sometimes can exhibit wave properties. On the
contrary, they arise quite naturally as soon as we accept the
fuzzy basis (meaning "being partly in...other states") of
microscale phenomena which implies, among other things, the
above-mentioned "self-interference."\\

How can we derive the equation that would incorporate these
essential features of microscale phenomena and, under certain
conditions, would yield the Hamilton-Jacobi equation of classical
mechanics? We start with the Hamilton-Jacobi equation (not
Newton's second law) because of its connection to the hidden
fuzziness in classical mechanics. We consider the simplest
classical object that would allow us to get the desired results
that will account for the two experimental facts mentioned
earlier.\\

We choose a free particle by setting $V = 0$ in Eq. (4). Our
problem is somewhat simplified now. We are looking for a linear
equation whose wave-like solution is simultaneously a solution of
the Hamilton-Jacobi equation. Since the mechanical phenomena
behave differently at microscales and macroscales, the linear
equation should contain a scale factor (that is to be
scale-dependent), such that in the limiting case corresponding to
the macroscopic value of this factor we get the nonlinear
Hamilton-Jacobi equation for a free particle.\\

A nonlinear equation admits a wave-like solution (for a complex
wave) if this equation is homogeneous of order $2$. Since Eq. (4)
does not satisfy this criterion, we cannot expect to find a wave
solution for the function $S$. However, this turns out to be a
blessing in disguise, because by employing a new variable in
place of the action $S$, we can both convert this equation into a
homogeneous (of order 2) equation (thus allowing for a wave-like
solution) and simultaneously introduce the scaling factor. It is
easy to show that there is one and only one transformation of
variables that would satisfy both conditions:
\begin{equation}
S=K ln\Psi
\end{equation}
where the scaling factor $K$ is to be found later.\\

Upon substitution of (7) in (4), with $V=0$, we obtain the
following homogeneous equation of the second order  with respect
to the new function $\Psi$:
\begin{equation}
K\Psi \frac{\partial \Psi}{\partial t}+\frac{K^2}{2m}(\nabla
\Psi)^2
\end{equation}
Equation (8) is easily solved by the separation of variables,
yielding
\begin{equation}
\Psi =C exp[-\frac{{at-(2m/a)^{1/2}\bf{a \bullet x}}}{K}]
\end{equation}
where the vector $\bf{a}$ of length $a$ is another constant of
integration. Since solution (9) must be a complex-valued wave, the
argument of $\Psi$ must satisfy two conditions:

i)it must be imaginary, and

ii)the factors at the variables $t$ and $\bf{x}$ must be the
frequency $\omega=2\pi \nu$ and the wave vector $\bf{k}$,
respectively.

This results in the following:
\begin{equation}
 K = -iB
\end{equation}
 and
 \begin{equation}
 a/B=\omega,~~ (2m/a)^{1/2} \bf{a/B}=\bf{k}
\end{equation}
where $B$ is a real-valued constant. Now the solution (9) is
\begin{equation}
\Psi=C exp[-i(\omega t -\bf{k \bullet x}]
\end{equation}

Since both functions $S$ and $\Psi$ are related by Eq. (7), we can
easily establish the connection between the kinematics parameters
of the particle and the respective parameters $\omega$ and
$\bf{k}$, which determine the wave-like solution of the
Hamilton-Jacobi equation for the new variable $\Psi$. According
to classical mechanics, $-\partial S/\partial t$ is the particle
energy $E_0$, and $\nabla S$ is the particle momentum $\bf{p}$. On
the other hand, these quantities can be expressed in terms of the
new variable $\Psi$ with the help of Eqs. (7) and (12), yielding
$E_0 = B\omega, B \bf{k} = \bf{p}$.\\

From these relations we see that for a free particle its energy
(momentum) is proportional to the frequency $\omega$ (wave vector
{\bf{k}} ) of the wave solution to the "scale-sensitive"
modification of the Hamilton-Jacobi equation. The constant $B$ is
found by invoking the experimental fact that $E_0 = h\nu =\hbar
\omega$ (where $h$ is Planck's constant). This implies $B = \hbar$
or $K = -i\hbar$, and as a byproduct, the de Broglie equation
${\bf{p}} = \hbar \bf{k}$. Inserting solution (12) into the
original nonlinear equation (8), we arrive at the dispersion
relation
\begin{equation}
 \omega=(\hbar/2m)k^2
\end{equation}

Now we can find the linear wave equation whose solution and the
resulting dispersion relation are given by Eqs.(12) and (13)
respectively. Using an elementary vector identity, we rewrite Eq.
(8):
\begin{equation}
 [\frac{\partial}{\partial t}-\frac{i\hbar}{2m}{\nabla}^2]\Psi -
 \frac{i \hbar}{2m \Psi}[div(\Psi \nabla \Psi)-2 \Psi
 {\nabla}^2\Psi]=0
\end{equation}\\

Equation (14) is the sum of the two parts, one linear and the
other nonlinear in $\Psi$. The solution (12) makes the nonlinear
part identically zero, and this solution, together with the
dispersion relation (13), must also satisfy the linear part of Eq.
(14). Therefore we have proven the following: If the wave-like
solution (12) satisfies Eq. (8), then it is necessary and
sufficient that it must be a solution of the following linear
partial differential equation, the Schroedinger equation:
\begin{equation}
[i \hbar \frac{\partial}{\partial
t}+\frac{{\hbar}^2}{2m}{\nabla}^2]\Psi=0
\end{equation}\\

Now we return to the variable $S$ according to $\Psi = exp
(iS/\hbar)$ and introduce the following dimensionless quantities:
time $\tau =t/t_0$, spatial coordinates ${\bf{R}} ={\bf{x}}/L_0$
, the parameter ${\bf{h}} = \hbar/S_0$ (which  we call the
Schroedinger number), and the dimensionless action ${\bf{S}} =
S/S_0$. Here, $S_0 = mL^2_0/t_0$, $L_0$ is the characteristic
length, and $t_0$ is the characteristic time. As a result, we
transform (15) into the following dimensionless equation:
\begin{equation}
\partial {\bf{S}}/\partial\tau + (1/2)(\nabla {\bf{S}})^2=(i {\bf{h}}/2)
\nabla^2 \bf{S}
\end{equation}\\

This equation is reduced to the classical Hamilton-Jacobi equation
(or, equivalently, the equation corresponding to the minimum
fuzziness)  if its right-hand side goes to 0. This is possible
only when the Schroedinger number $\bf{h}$ goes to 0. Therefore,
at least for a free particle, this number serves as a measure of
fuzziness of a microobject. Since $h$ is a fixed number, the limit
${\bf{h}} \rightarrow 0$ is possible only if $S_0 \rightarrow
\infty$, thus confirming our earlier assumption that action $S$
represents a measure of fuzziness of a microobject. For a free
particle this means that with the decrease of $S_0$ the fuzziness
of the particle increases.\\

Interestingly enough, the question of fuzziness (although not in
these terms) was addressed in one of the first six papers on
quantum mechanics written by Schroedinger \cite{ES}. He wrote,
"... the true laws of quantum mechanics do not consist of definite
rules for the single path, but in these laws the elements of the
whole manifold of paths of a system are bound together by
equations, so that apparently a certain reciprocal action exists
between the different paths."\\

It turns out that by using the same reasoning as for a free
particle we can easily derive the Schroedinger equation from the
Hamilton-Jacobi equation for  a piece-wise constant potential. If
we  replace in the resulting Schroedinger equation the function
$\Psi$ by $S$ according to (7), and introduce the dimensionless
variables used for a free particle we obtain
\begin{equation}
(\partial/\partial t){\bf{S}}+(1/2)(\nabla{\bf{S}})^2+ {\bf{U}}=
(i{\bf{h}}/2)\nabla^2{\bf{S}}
\end{equation}

where $\bf{U} = V/S_0$ is the dimensionless potential. Once again,
the Schroedinger number serves as the indicator of the respective
fuzziness, yielding the classical motion (a zero fuzziness) for
$\bf{h} \rightarrow 0.$\\

A more general case of a variable potential $V(\bf{x},t)$ cannot
be derived from the Hamilton-Jacobi equation with the help of the
technique used so far, since there are no monochromatic complex
wave solutions common to the nonlinear Hamilton-Jacobi equation
and the linear Schroedinger equation. Therefore  we postulate that
the Schroedinger equation describing a case of an arbitrary
potential $V(\bf{x},t)$ should have the same form as for a
piece-wise constant potential. This postulate is justified by the
fact that, apart from the experimental confirmations, in the
limiting case of a very small Schroedinger number, $\bf{h}
\rightarrow 0$ (minimum fuzziness), we recover the appropriate
classical Hamilton-Jacobi equation. In what follows we will
describe this process of recovering classical mechanics from
quantum mechanics (which we dubbed "defuzzification") in a
different fashion that will require a study of a physical meaning
of the function $\Psi$.

 \section{ FUZZINESS AND THE WAVE FUNCTION $\Psi$}
Earlier, by considering the Schroedinger number $\bf{h}$, we saw
that the action $S$ represents some measure of fuzziness.
Therefore, it is reasonable to expect that the function $\Psi =
exp (iS/\hbar)$ is also related to the measure of fuzziness. Since
the fuzziness is measured by real-valued quantities (degree of
membership, membership density), a possible candidate for such a
measure would be some function of various combinations of $\Psi$
and $\Psi^*.$ There is an infinite number of such combinations.
However, it is easy to demonstrate \cite{DB} that the
Schroedinger equation is equivalent to the two nonlinear coupled
equations with respect to the two real-valued functions
constructed out of $\Psi\Psi^*$ and $(\hbar/2i) ln (\Psi/\Psi^*).$
Therefore, our choice of all possible real-valued combinations is
reduced to only two functions. However, in the limiting
transition to the classical case,  $(\hbar/2i)ln (\Psi\Psi*)$ is
related to the classical velocity. Therefore we are left with only
one choice: $\Psi\Psi^*$.\\

The easiest way to find a physical meaning of $\Psi\Psi^*$ is to
consider some simple specific example that can be reduced to a
respective classical picture. To this end we consider a solution
of the Schroedinger equation for a free particle passing through a
Gaussian slit  \cite{RF}:
\begin{eqnarray}
\Psi =(\frac{m}{2\pi i\hbar})^{1/2} \frac{1}{(T+t+i\hbar
Tt/mb^2)^{1/2}} \nonumber \\
\times exp[\frac{im(v_0^2 T+x^2/t)}{2\hbar}+\frac{(m/\hbar
t)^2(x-v_0 t)}{2 im\hbar (1/T+1/t)-1/b^2}]
\end{eqnarray}
where $T$ is the initial moment of time, $t$ is any subsequent
moment of time, $b$ is the half-width of the slit, $v_0 = x/T$,
and $x_0$ is the coordinate of the center of the slit.\\

Using (18) we immediately find that $\Psi \Psi^*$ is
\begin{equation}
\Psi\Psi^* = \frac{1}{2p{\bf{h}}b^2}\frac{1}{\sqrt{(1+t/T
+{\bf{h}}^2}} exp[-\frac{{\bf{S}}}{1+t/T+{\bf{h}}^2}]
\end{equation}

where now ${\bf{S}} = (x -v_0 t/)b^2.~$  Executing the transition
to the case of a classical particle passing through an
infinitesimally narrow slit, we set both ${\bf{h}}\rightarrow 0$
and $b \rightarrow 0$. As a result, (19) will become the delta
function. Recalling that we define a classical mechanical
particle as a fuzzy entity with a delta-like membership density,
we arrive at the conclusion that the real-valued quantity $\Psi
\Psi^*$ can be identified as the membership density for a
microobject.\\

This allows one to ascribe to $\Psi \Psi^* dV$ the physical
meaning of the degree of membership of a microobject in an
infinitesimal volume $dV$ (cf. the analogous statement postulated
in Ref. \cite{BK2}). This in turn implies a nice geometrical
interpretation with the help of a generalization of Kosko's
multi-dimensional cube. Any fuzzy set A (in our case a fuzzy
state) is represented (see Fig. 1 for a two-dimensional cube) by
point A inside this cube. Following Kosko, we use the sum of the
projections of vector A onto the sides of the cube as the
cardinality measure.

Let us consider the following integral:
\begin{equation}
\int_{-\infty}^{\infty}\Psi \Psi^*dV = {\lim_{N \rightarrow
\infty}}\sum_{i=1}^{N}\Psi\Psi^*
\end{equation}

If this integral is bounded, then we can normalize it. As a
result, we can treat the right-hand side of (20) as the sum of
the projections of the "vector" $\int_{-\infty}^{\infty}\Psi
\Psi^*dV$ onto the sides $\Psi_i\Psi_i^*\Delta V_i$ of the
infinitely dimensional hypercube. This allows us to represent the
integral as the vertex A along the major diagonal of this
hypercube.\\

According to the subsethood theorem \cite{BK3} each side of the
hypercube represents the degree of membership of the microobject
(viewed as a deterministic fuzzy entity) in any given elemental
volume $dV_i$ built around a given spatial point $x_i$.
Respectively the relative membership of the microobject in two
different spatial points $x_i$ and $x_j$, that is,
$\Psi_i\Psi_i^*/\Psi_j\Psi_j^*$ is equal to the ratio of the
respective numbers of the successful outcomes in a series of
experiments aimed at locating the microobject (or rather its
part) at the respective elemental volumes. Hence we can conclude
that the membership density at a certain point is proportional to
the number of successful outcomes in repeated experiments aimed
at locating the fuzzy microobject at the respective elemental
volumes.\\

If the integral on the right-hand side of (20) is divergent, this
does not change our arguments, since $\Psi_i\Psi_i^*$ is a
measure of the successful outcomes in a series of experiments
that do not depend on the convergence of the integral. Thus we
see that the fuzziness, via its membership density, dictates the
number of successful outcomes in experiments aimed at locating
the fuzzy microobject. Continuing this line of thought we see
that any physical quantity associated with the fuzzy microobject
is not tied to a specific  spatial point. This indicates a need to
introduce a process of defuzzification with the help of the
membership density which would serve as the "weight" in this
process. Such defuzzification is different from what is usually
understood by this term, that is, a process of "driving" a fuzzy
point to a nearest vertex of a hypercube. Instead, we take the
degree of
 \begin{figure}
 \begin{center}
 \includegraphics[width=6cm, height=6cm]{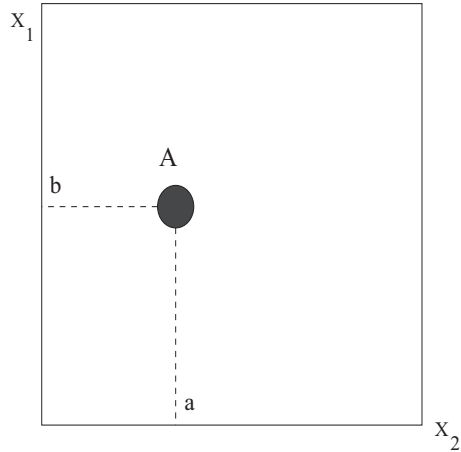}
 \caption{\small  Geometrical interpretation of fuzzy sets. The fuzzy
subset A is a point in the unit 2-cube with coordinates a and b.
The cube consists of all possible fuzzy subsets of two elements,
@$x_i$ and $x_j$.}
 \end{center}
 \end{figure}
membership $(\Psi\Psi^*)_i \Delta V_i$ at each vertex of the
infinite-dimensional hypercube and multiply it by the value of
the physical quantity at the respective point $x_i$. Summing over
all these products results in the averaged (defuzzified) value of
the quantity.

Thus, instead of averaging over the distribution of random
quantities, we introduce the defuzzification of deterministic
quantities. Mathematically both processes are identical, but
physically they are absolutely different. We do not need any more
the probabilistic interpretation of the wave function $\Psi$,
which implies that there is another, more detailed level of
description that would allow us to get rid of uncertainties
introduced by randomness. Now it is clear that, within the
framework of the fuzzy interpretation, we cannot get rid of the
uncertainties intrinsic to fuzziness (and not connected to
randomness). From this point of view quantum mechanics does not
need any hidden variable to improve its predictions. They are
precise within the framework of the fuzzy theory.\\

Moreover, since quantum mechanics is a linear theory, one can
speculate that according to the fuzzy approximation theorem
\cite{BK4} the linearity and fuzziness of quantum mechanics are
the best tools to approximate (with any degree of accuracy) any
macrosystem (linear or nonlinear). The linearity of quantum
mechanics is responsible for the uncertainty relations which are
present in any linear system. Therefore (as was demonstrated long
time ago \cite{LM}), these relations enter quantum mechanics even
before any concept of measurement. \\

Let us consider the membership density of a free microobject (a
progenitor of a classical free particle). It is obvious that
$\Psi\Psi^* = const$. This means that the relative degree of
membership for any two points in space is 1. In other words, the
free microobject is "everywhere,"  the same property that is
characteristic for a three-dimensional standing wave. In
particular, this example shows that the wave-particle duality is
not necessarily a duality but rather an expression of the fuzzy
nature of things quantum.\\

In fact, we can even go that far as to claim that the
complementarity principle is a product of a compromise between
the requirements of the bivalent logic and the results of quantum
experiments. Within the framework of the fuzzy approach there is
no need to require complementarity, since the logic of a fuzzy
microobject transcends the description of its properties in terms
of either/or and, as a result, is much more complete, probably
the most complete description under the given experimental
results.\\

It turns out that the membership density has something more to
offer than simply a degree to which a fuzzy microobject has a
membership in a certain elemental volume dV. In fact, using the
expansion of the wave amplitude (we could call it "fuzziness
amplitude") in its orthonormal eigenfunctions $\Psi$ and assuming
that the integral in (20) is bounded, we write the well-known
expression
\begin{equation}
\int_{-\infty}^{\infty}\Psi\Psi^* dV = \sum_{k=1}^{\infty}a_k
a_k^*=1
\end{equation}

Equation (21) allows a very simple geometric interpretation with
the help of a $(N - 1)$-dimensional simplex. A fuzzy state $\Psi$
is represented as a point $A$ at the boundary of this simplex.
(Figure 2 shows this for a one-dimensional simplex, $k = 1,2.$)
Its projections onto the respective axes correspond to the values
$a_ka_k^*$.\\

Now applying the subsethood theorem \cite{BK2}, we interpret the
values of $a_ka_k^*$ as the degree to which the state A is
contained in a particular eigenstate $k$. Using Fig. 2 we can
clearly see that $A \cap B = B, A \cap C = C$. Moreover, the same
figure shows that the lengths of projections of A onto the
respective axes (namely, OA and OC) are nothing but the
cardinality sizes $M(A \cap B) = a_1a_1^*$ and $M(A \cap C) =
a_2a_2^*$. On the other hand, the cardinality size of A is $M(A)
= 1$. Therefore, the respective subsethood measures are $S(A,B) =
a_1a_1^*/1$ and $S(A,C) = a_2a_2^*/1.$ At the same time, both of
these measures provide a number of  detections (successful
outcomes) of the respective states $k = 1$ or $k = 2$ in the
repeated experiments.\\

Our discussions is applicable to a particular case of a state A
described by a wave (fuzziness) amplitude $\Psi$ corresponding to
a pure state. However it is general enough to describe a mixed
state characterized by the density matrix $\rho(x',x)$. The
integral of $\rho(x',x)$ over all $x's$ yields the sum
$\sum_{k=1}^{\infty}a_{kk'}$ which is the generalization of a
measure of containment of the fuzzy state A in the discrete
states k.\\

\begin{figure}
 \begin{center}
 \includegraphics[width=6cm, height=6cm]{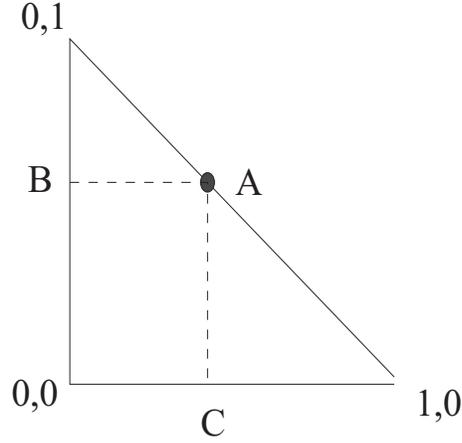}
 \caption{\small Representation of a quantum mechanical state $A$ as a point
  in a one-dimensional fuzzy simplex}
 \end{center}
 \end{figure}

By preparing a certain state, which is now understood to be a
fuzzy entity, we fix the frequencies of the experimental
realizations of this fuzzy state in its substates $k$. If the
fuzzy state A undergoes a continuous change, which corresponds in
Fig. 2 to motion of point A along the hypotenuse, then its
subsethood in any state $k$ changes. This implies the following:
if the eigenfunctions of a fuzzy set stay the same, the degree to
which the respective eigenstates represent the fuzzy state
varies. The variation can occur continuously despite the fact
that the eigenstates are discrete.\\

This indicates an interesting possibility that quantum mechanics
is not necessarily tied to the Hilbert space. Such a possibility
was mentioned long ago by von Neumann \cite{vN} and recently was
addressed by Wulfman \cite{CW}. One of the hypothetical
applications of this idea is to use quantum systems as an
infinite continuum state machine in a fashion that is typical for
a fuzzy system: small continuous changes in the input from some
"ugly" nonlinear system will result in small changes at the output
of the quantum system which in turn can be correlated with the
input to produce the desired result.\\

Concluding our introduction to a connection between fuzziness and
quantum mechanics, we prove a statement that can be viewed as a
generalized Ehrenfest theorem. We will demonstrate that
defuzzification of the Schroedinger equation (with the help of the
membership density $\Psi\Psi^*$) yields the Hamilton-Jacobi
equation. This will provide  $\it {aposteriori}$ derivation of the
Schroedinger equation for an arbitrary potential $V({\bf{x}},t)$.
We assume that the fuzzy amplitude $\Psi \rightarrow 0$ as
${\bf{x}}\rightarrow \infty$ and rewrite the Schroedinger equation
as follows:
\begin{equation}
\frac{\hbar}{i}\frac{\partial  ln\Psi}{\partial
t}+\frac{\hbar^2}{2m}{(\nabla ln \Psi)}^2 +V = 0
\end{equation}
Integrating (22) with the weight $\Psi\Psi^*$ (i.e.,
"defuzzifying" it), we obtain
\begin{equation}
\int \Psi\Psi^*[\frac{\hbar}{i}\frac{\partial  ln\Psi}{\partial
t}+\frac{\hbar^2}{2m}{(\nabla ln \Psi)}^2 +V]d^3x = 0
\end{equation}
Integrating the second term by parts and taking into account that
the resulting surface integral vanishes because $\Psi \rightarrow
0$ at infinity, we obtain the following equation:
\begin{equation}
\langle \frac{\partial S}{\partial t}\rangle +\frac{1}{2m}\langle
\nabla S \nabla S^*\rangle +\langle V\rangle =0
\end{equation}

where $\langle \rangle$ denote defuzzification with the weight
$\Psi\Psi^*$, and $S = (\hbar/i) ln\Psi$. This equation is
analogous to the classical Hamilton-Jacobi equation (4).\\

The generalized Ehrenfest theorem shows that the classical
description is true only on a coarse scale generated by the
process of "defuzzification," or measurement. The "classical
measurement" corresponds to the introduction of a non-quantum
concept of the potential $V({\bf{x}},t)$ serving as a shorthand
for the description of a process of interaction of a microobject
(truly quantum object) with a multitude of other microobjects.
This process destroys a pure fuzzy state (a constant fuzziness
density) of a free quantum "particle."

Paraphrasing Peres, \cite{P} we can say that a classical
description is the result of our "sloppiness," which destroys the
fuzzy character of the underlying quantum mechanical phenomena.
This means that, in contradistinction to Peres, we consider these
phenomena "fuzzy" in a sense that the respective membership
distribution in quantum mechanics does not have a very sharp peak,
characteristic of a classical mechanical phenomena. Note that we
exclude from our consideration the problem of the classical
chaos,assuming that our repeated experiments are carried out
under the absolutely identical conditions.

\section{CONCLUSION}

This work represents a continuation of our previous effort
\cite{GC1} to understand quantum mechanics in terms of the fuzzy
logic paradigm. We regard reality as intrinsically fuzzy. In
spatial terms this is often called nonlocality. Reality is
nonlocal temporarily as well, which means that any microobject has
membership (albeit to a different degree) in both the future and
the past. In this sense one might define the present as the time
average over the membership density. A measurement is defined as a
continuous process of defuzzification whose final stage,
detection, is inevitably accompanied by a dramatic loss of
information through the emergence of locality, or crispness, in
fuzzy logic terms.\\

We have attempted to provide a description of quantum mechanics
in terms of a deterministic fuzziness. It is understood that this
attempt is inevitably incomplete and has many features that can
be improved, extended, or corrected. However, we hope that this
work will inspire others to start looking at the quantum
phenomena through "fuzzy" eyes, and perhaps something practical
(apart from removing wave-particle duality and complementarity
mysteries) will come out of this.\\

{\bf{Acknowledgment}}

One of the authors (AG) wishes to thank V. Panico for very long
and very illuminating discussions, which helped to shape this
work, and for reading the manuscript. HJC's work was supported by
the Air Force Office of Scientific Research

\end{document}